\documentclass[nato,noid,numreferences]{crckbked}

\usepackage{graphicx}
\usepackage{epsf}
\newcommand{\tfrac}[2]{{\textstyle\frac{#1}{#2}}}

\begin{document}

\begin{opening}
\title{Charges, Monopoles and Gauge Invariance}

\author{R Horan}
\author{M Lavelle}
\author{D McMullan}
\institute{Department of Mathematics and Statistics\\
           University of Plymouth\\Plymouth PL4 8AA, UK}

\author{and}
\institute{}
\author{A Khvedelidze} 
\institute{Laboratory of Information Technologies\\
Joint Institute for Nuclear Research\\141980 Dubna, Russia}

\begin{abstract}
A systematic approach to the description of gauge invariant
charges is presented and applied to the construction of both the
static colour charge configuration in QCD and the monopole
solution in pure SU(2). The gauge invariant non-abelian monopole
offers a new style of order parameter for monopole condensation.
\end{abstract}
\renewcommand{\thefootnote}{\fnsymbol{footnote}}
\footnotetext[0]{Talk given by D McMullan at the NATO workshop on
``Confinement, Topology, and other Non-Perturbative Aspects of
QCD", Star\'a Lesn\'a, Slovakia, January 21-27, 2002.}
\renewcommand{\thefootnote}{\arabic{footnote}}
\end{opening}

\section{Introduction}
It is a fact of life, and one that has been exploited again and
again at this meeting, that even gauge invariant objects are more
transparent in a specific gauge. For example, the static
interquark potential, which has  a well known gauge invariant
definition, is often best treated in Coulomb gauge (see, for
example, \cite{Gribov:1977mi,Drell:1981gu} and more
recently~\cite{Greensite:2001nx}) as this gauge is, in some way,
closely adapted to that physical system. Usually such a choice of
gauge is a simple pragmatic decision based on the simplicity of
the resulting calculation. However, what we'd like to argue  is
that the connection between such an adapted gauge and the correct
gauge invariant description, often goes much
deeper~\cite{Lavelle:1997ty}. We will see that such appropriate
gauge fixings, once recognised, can lead to an understanding of
the dominant contribution to the gauge invariant description of
the relevant physical degrees of freedom. For monopoles and
vortices, where gauge invariant formulation do not yet exist,
understanding this route from gauge fixing to gauge invariance is
of central importance if we are ever to understand fully their
roles in the non-perturbative structure of QCD.

\section{What is a static quark?}
In order to map out the connection between static quarks and the
Coulomb gauge, we need to make precise just what we mean by saying
that a field $\Psi$ describes a static quark. First, to capture
the requirement that quarks carry colour, the field $\Psi$
\emph{cannot} be a coloured singlet under the global gauge
transformation although it \emph{must} be gauge invariant under
the local gauge transformations. This tells us straight away that
it cannot be identified with the matter field $\psi$ that enters
directly into the formulation of the theory since under a gauge
transformation we have $\psi\to\psi^U=U^{-1}\psi$. What we must
have is that $\Psi=h^{-1}\psi$, for some field dependent
configuration $h^{-1}$, where under a gauge transformation we have
\begin{equation}\label{gid}
  h^{-1}\to (h^{-1})^{U}=h^{-1}U\,.
\end{equation}
We call $h^{-1}$ a dressing for the charge. It incorporates the
cloud of fields around any charge.

To impose the static condition that $\partial_0\Psi=0$, we need to
realise that static means\footnote{Alternatively, rather than
infinite mass, we can consider asymptotic fields that are static.
See the discussion in Ref.\cite{Horan:1999ba}} infinite mass and
hence we can exploit the dynamical simplification that comes from
the heavy quark effective theory. That is, we can use the
equations of motion that the matter field is covariantly constant:
$(\partial_0+gA_0)\psi=0$. Using this, and the condition that
$\Psi$ is static, it is easy to see that the dressing must satisfy
the static dressing equation
\begin{equation}\label{sd}
  \partial_0h^{-1}=gh^{-1}A_0\,.
\end{equation}

Equations (\ref{gid}) and (\ref{sd}) are the fundamental
conditions the dressing must satisfy in order to construct a
static charge. Explicit solutions to them can be found in
QED~\cite{Horan:1998im,Bagan:1999jf}, and perturbative solutions
to them can be constructed in QCD~\cite{Lavelle:1997ty}. The point
to note here is that the resulting dressing has \emph{structure}.
The charged field factorises into the product of two separately
gauge invariant terms
\begin{equation}\label{df}
    \Psi=h^{-1}\psi=\left(\tilde\mathrm{{T}}\,\mathrm{e}^{-K}\right)\mathrm{e}^{-\chi}\psi
\end{equation}
The bracketed term involves an anti-time ordering while the rest
of the expression is local in time but non-local in space. We call
$\mathrm{e}^{-\chi}$ the minimal part of the dressing as it is
essential for the overall gauge invariance of the charge.
Additional terms, such as $\mathrm{e}^{-K}$  in (\ref{df}), are
not expected from the overall requirement of gauge invariance.
Rather, they are needed to ensure the correct dynamical properties
of the charge.

The significance of this factorisation can be seen in either the
infra-red properties of the fields or in the forces  between two
such charges. It emerges~\cite{Bagan:1999jk} that in QED the
minimal part of the dressing is responsible for controlling the
soft infra-red structure of the theory, while the additional part
deals with the phase divergences. In terms of
forces~\cite{Lavelle:1998dv,Bagan:2000nc,Bagan:2001wj}, the
minimal part gives the anti-screening contribution to the
inter-quark potential, while the additional term is needed for the
lesser screening forces. Given the dominance of anti-screening
over screening, we see that the minimal part of the dressing is
capturing the dominant glue content of a static charge.

There are many important and interesting properties of these
fields, but the key thing to note here is that the minimal part of
the dressing becomes the identity in Coulomb gauge. This important
fact is most easily seen in QED where
\begin{equation}\label{qedmin}
\chi(x)\propto\int \!d^3z
\phi_i(x-z)A_i(z)=\frac{\partial_iA_i}{\nabla^2}(x)\,,
\end{equation}
and $\phi_i$ is the classical Coulombic electric field of a static
charge.

We see that Coulomb gauge is the unique gauge that trivialises the
dominant part of the static dressing. This now makes precise the
sense in which  the Coulomb gauge is adapted to the description of
static charges. Knowing this connection can, in  turn, provide an
efficient means for calculating the minimal part of the dressing
which is, as we've seen, essential for a gauge invariant
description of the static charge.

This connection between gauge invariance and gauge fixing can be
generalised to moving
charges~\cite{Bagan:1997su,Horan:1998im,Bagan:1999jf} and it can
also be given a simple geometric and hence global
interpretation~\cite{Lavelle:1997ty}. The conclusion from such an
analysis is that in QCD there is a global obstruction to the
construction of a static coloured charge, and that this is how
confinement is seen in this approach.

\section{Gauge invariant monopoles}
We now want to consider a pure non-abelian gauge theory and show
how a monopole creation operator can be constructed. The are
several reasons for wanting to do this, the most immediate being
to allow for the construction of new order parameters with which
the dual superconductor account of confinement can be tested. It
is also, as we'll see, an interesting theoretical study of the
relation between classical solutions and quantum configurations in
gauge theories.

To motivate our approach, we note that in Dirac's original account
of the construction of electric charges~\cite{Dirac:1955uv}, he
arrived at the minimal, abelian, static dressing (\ref{qedmin}) by
noting that its commutator with the electric field operator
generated the Coulombic electric field expected from  a static
charge. As we will see, a similar argument can be applied to
monopoles.

Let us start again in the abelian theory. Suppose that $f_i(x)$ is
the classical Dirac monopole potential~\cite{Dirac:1948um}. Then
it is straightforward to see that the operator
\begin{equation}\label{dm}
    M=\exp\left(i\int \!dz\,f_i(z)E_i(z)\right)\,,
\end{equation}
is gauge invariant\footnote{There is an interesting subtlety here
since different gauge related potentials $f_i$ will yield
different, but weakly equivalent monopole operators. Thus the
ability to move the Dirac string is seen in the weak equivalence
of the construction. } and its equal-time commutator with the
potential is
\begin{equation}\label{et}
    [A_i(x),M]=f_i(x)M\,.
\end{equation}
So we can interpret $M$ as a monopole creation operator for the
pure abelian theory. Although this operator allows us to rederive
many of the important properties of monopoles, the singularity of
the potential $f_i$ makes this an artificial construction in QED.
This reflects the fact that we do not expect  monopoles in such an
abelian theory. However, in  non-abelian theories regular monopole
solutions are known to exist when spontaneous symmetry breaking
occurs, and they are conjectured to exist and to play an important
role in pure gauge theories (see for example,
Ref~\cite{'tHooft:1999au}). With this in mind, we now investigate
how  (\ref{dm}) can be generalised to the non-abelian theory.

The naive extension of this simple construction to a non-abelian
theory, where we replace the abelian potential $f_i$ by a
non-abelian one $f_i^a\tau^a$ and the electric field $E_i$ by its
chromo-electric generalisation $E_i^a\tau^a$, runs into two
immediate problems. The first is to decide on how to generalise
the classical  Dirac monopole configuration. The second is
maintaining gauge invariance since, as is well known, the
chromo-electric field is not gauge invariant.

If we now specialise to a pure SU(2) gauge theory, then there is a
natural candidate for a monopole configuration first written down
by Wu and Yang~\cite{wuyang}:
\begin{equation}\label{wy}
    f_i^a=\tfrac1g\varepsilon_{aib}\frac{x^b}{r^2}\,.
\end{equation}
This is a solution to the classical Yang-Mills equations of motion
which, through a singular gauge transformation, can be related to
the abelian monopole configuration. It should be noted, though,
that it is an unstable
solution~\cite{Yoneya:1977yi,Brandt:1979kk}. How this is modified
or reflected in the quantum theory is, however, unknown.

The gauge non-invariance of the chromo-electric field seems a much
more serious obstacle to the construction of a non-abelian
generalisation of (\ref{dm}). Extending the dressing technique
used in the description of a static charge, we will solve this
problem by dressing the chromo-electric field
\begin{equation}\label{ced}
    E_i^a\tau^a\to  \tilde{E}_i^a\tau^a=h^{-1} E_i^a\tau^ah\,,
\end{equation}
where the dressing transforms as in (\ref{gid}) so that
$\tilde{E}_i^a$ is now gauge invariant. The monopole creation
operator generalising (\ref{dm}) is then
\begin{equation}\label{su2dm}
    M=\exp\left(i\int \!dz\,f_i^a(z)\tilde{E}_i^a(z)\right)\,.
\end{equation}

Given that we want to describe a static monopole, i.e.,  it should
not generate a chromo-electric field, we require $[E_i^a(x),M]=0$.
This will follow if $h=h[E]$, i.e.,  the dressing must solely
depend on the chromo-electric field. As such, we cannot simply use
the dressing constructed in (\ref{df}) to describe a static quark.
To proceed, we recall the close relation between monopoles and
abelian gauge fixing~\cite{'tHooft:1999au}. Following  our method
for describing the minimally dressed static quark, we will exploit
this adapted class of gauge fixings to construct a gauge invariant
chromo-electric field and hence the dressing needed
in~(\ref{ced}).

Gauge fixing in the chromo-electric sector is, as far as we know,
not well studied.  The interesting point here is that it is not
possible to fully fix the gauge. In terms of constraints, one can
easily see that it is impossible to construct a complete  second
class set out of Gauss' law and functions of just the
chromo-electric field. However, second class subsets can be found
that are valid on regions of the phase space. To see how this
works, consider the simple chromo-electric gauge
\begin{equation}\label{abeliang}
E_3^1=E_3^2=0\,.
\end{equation}
These, along with the components $D_iE_i^1$ and $D_iE_i^2$ of
Gauss' law, form a second class set of constraints as long as
$E_3^3\ne0$. To implement this reduction then, we should restrict
ourselves to the regions in phase space  where either $E_3^3>0$ or
$E_3^3<0$. If $E_3^3=0$, then we can either take it and one of the
components in (\ref{abeliang}) as our gauge, or we can choose
another  gauge by looking at, say, the $E_1^a$ components of the
chromo-electric field. In this way, through a patching process, we
can implement a chromo-electric gauge fixing that is only ill
defined on configurations which have zero field strength. We do
not yet fully understand the effect of such instanton
configurations on our monopole construction, so for the moment we
will neglect them and, for simplicity, just consider the gauge
(\ref{abeliang}) in the region $E_3^3>0$.

Having settled on a gauge that we know is adapted, or at least
sympathetic, to the non-abelian monopole configuration, we now
have to find the dressing needed in (\ref{ced}) by rotating our
fields into the gauge fixed configuration. For a configuration
space gauge fixing, such as the  Coulomb gauge, we were guaranteed
that the resulting dressing would at least locally satisfy the
fundamental relation (\ref{gid}). The incompleteness of the
chromo-electric gauge fixing, though, means that a little more
work is needed to get the correct transformation properties of the
dressing.

In terms of the dressed fields (\ref{ced}), we need to solve the
equations $\tilde{E}_3^1=\tilde{E}_3^2=0$. Now
\begin{equation}\label{etp}
\tilde{E}_3^a=-2E_3^b\mathrm{tr}\,(\tau^a h^{-1}\tau^b h)=E_3^b
R_{ab}\,
\end{equation}
where $R_{ab}$ is a rotation matrix. Hence we wish to solve $E_3^b
R_{1b}=0$ and $E_3^b R_{2b}=0$.

These two equations are simple vector equations and can be
immediately solved as follows. Take
 \begin{equation}\label{r1b}
R_{1b}=\varepsilon_{bcd}\hat{E}_3^c\hat{\lambda}^d
\end{equation}
 where
 \begin{equation}\label{ehat}
    \hat{E}_3^c=\frac{E^c_3}{\sqrt{E_3^eE_3^e}}
\end{equation}
and $\hat{\lambda}^d$ is, for the moment, an arbitrary unit
vector. Then
\begin{equation}\label{r2b}
    R_{2b}=\varepsilon_{bcd}\hat{E}_3^cR_{1d}
\end{equation}
and
\begin{equation}\label{r3b}
    R_{3b}=\varepsilon_{bcd}R_{1c}R_{1d}=\hat{E}_3^b\,.
\end{equation}
These allow us to construct the rotation matrix and check gauge
invariance of the resulting dressed chromo-electric field.

For the third colour component gauge invariance is immediate since
\begin{equation}\label{e3g}
    \tilde{E}_i^3=\frac{E_i^bE_3^b}{\sqrt{E_3^cE_3^c}}\,.
\end{equation}
We further note that $\tilde{E}_3^3=\sqrt{E_3^aE_3^a}$, which
(with our restriction that~$E_3^3>0$) is just $E_3^3$ in the gauge
(\ref{abeliang}).

 For the other components of the dressed chromo-electric field, though, gauge invariance
 can only be ensured through a good choice  of $\hat{\lambda}$. From the definition
 (\ref{etp}) we have
\begin{equation}\label{e1g}
    \tilde{E}_i^1=E_i^bR_{1b}=\varepsilon_{bac}E_i^b\hat{E}_3^a\hat{\lambda}^c\,,
\end{equation}
and
\begin{equation}\label{e2g}
    \tilde{E}_i^2=E_i^bR_{2b}=E_i^b\hat{E}_3^b\hat{E}_3^a\hat{\lambda}^a-
    E_i^b\hat{\lambda}^b\hat{E}_3^a\hat{E}_3^a\,.
\end{equation}
Gauge invariance will follow if $\hat{\lambda}^c$ is proportional
to $E^c$. However, from (\ref{r1b}), we also need $\hat{\lambda}$
orthogonal to $\hat{E}_3^a$. There are various ways to satisfy
these conditions for gauge invariance.
 For example, we could take
\begin{equation}\label{lambda}
    \hat{\lambda}^c=\frac{x_iE_i^c}{\sqrt{x_jE_j^dx_kE_k^d}}
\end{equation}

In summary, we have seen in this section  how to generalise the
abelian monopole creation operator~(\ref{dm}) to yield a gauge
invariant monopole operator~(\ref{su2dm}). This was done by
construction a chromo-electric dressing adapted to the
chromo-electric gauge fixing~(\ref{abeliang}).

\section{Conclusions}
An important contribution to the dressing approach to gauge
invariance is the recognition that there are special adapted
gauges that have a particular significance for the description of
both chromo-electric and magnetic charges in non-abelian gauge
theories. For electric charges in both QED and QCD, these adapted
gauges followed naturally from a more fundamental dressing
equation. Solving  that equation factorises the dressing into a
dominant anti-screening term (that controlled the soft infra-red
sector) and an additional screening term. For a specific dynamical
configuration for the charges, the adapted gauge trivialises the
dominant part of the dressing. However, it should be stressed that
in a scattering situation where charges with differing  momentum
must be dressed differently, there is no gauge in which all the
different dressings so simplify.

In pure SU(2) theory we have seen how to go from chromo-electric
gauge fixing to a gauge invariant monopole creation operator. As
yet there is no analogous dynamical approach to this dressing. It
is hoped, though, that through our recognition of the adapted
gauge to this system we have also captured the dominant monopole
configuration. This will allow us to further probe the role of
monopoles in confinement.

\end{document}